# Chemical bonds and spin state splittings in spin crossover complexes. A DFT and QTAIM analysis

**Hauke Paulsen**[1,*], **Juliusz A. Wolny**[1,2], **Alfred X. Trautwein**[1]

[1]Institut für Physik, Universität zu Lübeck, D-23538 Lübeck, Germany

[2]Wydział Chemii, Uniwersytet Wrocławski, 50-370, Wrocław, Poland

**Summary.** Density functional theory (DFT) calculations have been performed for the high-spin (HS) and low-spin (LS) isomers of a series of iron(II) spin crossover complexes with nitrogen ligands. The calculated charge densities have been analyzed in the framework of the quantum theory of atoms in molecules (QTAIM). For a number of iron(II) complexes with substituted tris(pyrazolyl) ligands the energy difference between HS and LS isomers, the spin state splitting, has been decomposed into atomic contributions in order to rationalize changes of the spin state splitting due to substituent effects.

____

*Corresponding author. E-mail: paulsen@physik.uni-luebeck.de

**Introduction**

Transition metal complexes that exhibit a temperature or pressure dependent reversible crossover from a low-spin (LS) to a high-spin (HS) state have been investigated since the beginning of the last century [1]. These spin crossover (SCO) complexes are regarded as very promising materials (e.g. for display or memory devices) since it was discovered that the spin crossover can be induced by irradiation with light [2,3]. In order to obtain a transition metal complex that exhibits a spin crossover, a delicate balance must be kept of the metal-ligand bond strength on one side and the electron-electron repulsion of the metal valence electrons on the other side. A quantitative analysis of the spin crossover phenomenon on the microscopic scale thus requires a quantitative analysis of the metal ligand bonds.

Most commonly, the molar HS fraction as a function of pressure and temperature, $\gamma_{HS}(p,T)$, is used as order parameter of the spin transition. For many solid samples $\gamma_{HS}$ exhibits features like abruptness or thermal hysteresis that are due to the cooperativity of many SCO molecules. However, in cases where the transition is gradual and without hysteresis, it can be described by a simple model [4] that is restricted to isolated complexes and requires only the knowledge of the difference $\Delta G = G^{HS} - G^{LS}$ between the Gibbs free energy of the HS and the LS isomers (it is not necessary to take the free enthalpy, since the term $pV$ is of the order of less than 1 J mol$^{-1}$ at ambient pressure and temperature). In this model the temperature dependence of the molar HS fraction $\gamma_{HS}$ can be written as

$$\gamma_{HS} = 1 / [1+\exp(\Delta G/k_B T)]. \qquad (1)$$

The free energy difference is a sum of three terms, the spin state splitting $E_S$, the vibrational energy difference $\Delta E_{vib}$ and the entropy difference multiplied by the temperature:

$$\Delta G(T) = E_S + \Delta E_{vib}(T) + T\Delta S(T). \qquad (2)$$

Only the latter two terms on the right side of Eq. (2) are temperature dependent, whereas the spin state splitting, which is in the order of a few thousand Kelvin, is in good approximation temperature independent. With the help of Eqs. (1) and (2) the transition temperature $T_{1/2}$, that is implicitly defined by $\gamma_{HS}(T_{1/2}) = 1/2$, can be written as

$$T_{1/2} = [E_S + \Delta E_{vib}(T_{1/2})] / \Delta S(T_{1/2}). \qquad (3)$$

Neglecting the vibrational energy difference $\Delta E_{vib}(T)$, which is rather small in comparison with the spin state splitting $E_S$ and in the range of the error margin of $E_S$ [5], Eq. (3) simplifies to

$$T_{1/2} = E_S / \Delta S_{1/2}, \qquad (4)$$

where $\Delta S_{1/2}$ denotes entropy difference at the transition temperature. From the two quantities on the right side of Eq. (4) it seems to be the spin state splitting $E_S$, which is most sensitive upon small variations of the SCO complexes, such as substitutions on the ligands or change of counterions. Considering a given class of similar SCO complexes, one may as a crude approximation take the entropy difference at the transition temperature as a constant proportionality factor and write simply

$$T_{1/2} \sim E_S. \qquad (5)$$

The crude approximations made here are justified insofar as the spin state splitting can be calculated anyway only with a very limited accuracy. The spin state splitting is calculated as the difference of energy between the HS and the LS isomer. The absolute values of these energies are more than five orders of magnitude larger than the difference. Therefore, any calculated value for $E_S$ has to rely on a very fortunate cancellation of errors [6-8].

A common approach to rationalize the spin state splitting is based of ligand field theory. This will be outlined here for the special but important case of complexes that have a central iron(II) ion with nearly octahedral coordination. By far the largest number of SCO complexes that are currently known belong to this class. The electronic ground states of the LS and HS isomers are usually denoted by $^1A_{1g}(t_{2g}^6)$ and $^5T_{2g}(t_{2g}^4 e_g^2)$, respectively. These terms refer to the irreducible representations of the $O_h$ group. In the ligand field model [9] the electronic state of the diamagnetic LS isomer is characterized by three fully occupied $3d$ orbitals that form a $t_{2g}$ representation ($d_{xy}$, $d_{xz}$, $d_{yz}$) and by two empty 3$d$ orbitals that form a $e_g$ representation ($d_{x^2-y^2}$, $d_{z^2}$). In the paramagnetic HS isomer, five electrons belonging to the majority spin are distributed over all five 3$d$ orbitals according to Hund's rule, the sixth electron that belongs to the minority spin enters a $t_{2g}$ orbital and hence the spin multiplicity 2$S$+1 equals 5. In molecular orbital (MO) theory the linear combinations of 3$d$ atomic orbitals centred on the metal ion and 2$p$ orbitals centred at the ligand atoms are formed. In this picture the metal 3$d$ MOs are those linear combinations for which the contributions of metal atomic orbitals are predominant. In a perfectly octahedral symmetry the $e_g$ metal MOs are anti-bonding and the $t_{2g}$ MOs non-bonding. Consequently, the metal ligand bond length, which for SCO complexes with [Fe(II)N$_6$] core is typically about 1.95 to 2.05 Å in the LS isomer, increases by about 10 % upon crossover to the HS isomer. The increase of bond lengths is accompanied by a decrease of the ligand field strength 10$Dq$. In case of an octahedrally coordinated metal ion with six 3$d$ electrons, the ligand field strength equals

$$10Dq = 2\,[\varepsilon(e_g) - \varepsilon(t_{2g})], \qquad (6)$$

where $\varepsilon(e_g)$ and $\varepsilon(t_{2g})$ describe the energies of the respective group of orbitals. The electron-electron repulsion $\Pi$ between the metal 3$d$ electrons depends far less on the metal ligand bond length and is often assumed to be constant with respect to the spin crossover. The energy needed for a vertical or Franck-Condon excitation from the LS ($^1A_{1g}$) to the HS ($^5T_{2g}$) state at constant LS geometry can be expressed by

$$\Delta E_{\text{HS-LS}}^{(\text{vert})} = 10Dq^{(\text{LS})} - \Pi. \qquad (6)$$

The superscript for the ligand field strength $10Dq^{(\text{LS})}$ indicates that this value refers to the geometry of the LS isomer. In order to calculate the energy needed for an adiabatic excitation from the LS isomer to the HS isomer, that is the spin state splitting energy $E_S$, one has to take into account the energy that is necessary for the rearrangement of the molecular geometry. The simplest picture that can be used to describe the geometry change upon spin crossover is the isotropic "breathing" of the [FeN$_6$] core. If the energy change can be described by one harmonic potential with the same force constant $k_{\text{LS}}$ for all six metal ligand bonds, the spin state splitting energy can be written as

$$E_S = 10Dq^{(\text{LS})} - \Pi + 6\, k_{\text{LS}}\, (r_{\text{HS}} - r_{\text{LS}})^2/2, \qquad (7)$$

where $r_{\text{HS}}$ and $r_{\text{LS}}$ denote the metal ligand bond lengths for the respective isomers. Regarding a vertical excitation of a HS isomer followed by an adiabatic geometry relaxation leads to an equivalent expression for the spin state splitting energy,

$$E_S = 10Dq^{(\text{HS})} - \Pi - 6\, k_{\text{HS}}\, (r_{\text{HS}} - r_{\text{LS}})^2/2. \qquad (7)$$

Obviously, not all parameters on the right sides of Eqs. (6) and (7) can be independent. However, due to the approximate character of the model it will be hardly reasonable to combine the two equations in such a way that one of the parameters could be eliminated. Instead it will be more instructive to sum up both equations in order to get an expression

$$E_S = (10Dq^{(HS)} + 10Dq^{(LS)})/2 - \Pi + 3(k_{HS} + k_{LS})(r_{HS} - r_{LS})^2/2 \qquad (8)$$

for the spin state splitting that is symmetric with respect to the spin states. All but one parameter on the right side of Eq. (8), the ligand field strength, the force constant and the metal ligand bond length for both isomers, can in principle be determined experimentally. Only the electron-electron repulsion $\Pi$ is not directly accessible. Assuming that $\Pi$ is to a good accuracy constant for similar SCO complexes, this parameter is cancelled if trends of the spin state splitting are calculated, like for instance

$$\Delta E_S = E_S^{(\alpha)} - E_S^{(\beta)}, \qquad (9)$$

where the Greek letters in brackets denote two different complexes. In this way ligand field theory allows to determine differences of spin state splitting if the ligand field strength $10Dq$, the force constant $k$ and the bond length $r$ is known for the respective spin states and complexes. Although this ligand field model gives qualitatively correct and clear picture, it does have only limited predictive power since $10Dq$, $k$ and $r$ have to be taken from experiments and the approximations made above limit the accuracy of the calculated spin state splitting.

An alternative to the ligand field model is given by modern electronic structure calculations based on density functional theory (DFT). These methods do not give such a direct and easy to grasp physical insight as provided by ligand field theory. Quite on the contrary, DFT methods may in this context be regarded as 'black box'. The advantage of DFT methods is that these methods allow in principle to calculate the spin state splitting accurately. Some progress has been made recently concerning new density functionals [10,11] and the inclusion of solid state effects [12]. The most reliable approach seems to be to calculate the trend of the spin state splitting $E_S$ for a given class of SCO complexes. The calculated trends are roughly independent on the choice of the density functional and consistent with experimental data [13].

Comparing the advantages of ligand field theory and density functional theory, the first approach gives clear physical insight whereas the latter has a reasonable predictive power. In order to combine the advantages of both approaches one may try to interpret the charge density obtained by DFT calculations in the framework of the *quantum theory of atoms in molecules* (QTAIM) described in detail in another contribution to this volume [14,15]. This theory allows finding in a rigorous way correspondences between the calculated charge densities and chemical concepts, like additivity schemes and transferability of the properties of functional groups.

< Fig. 1 >

In the present study iron(II) complexes with substituted tris(pyrazolyl) ligands have been investigated by a variety of different DFT methods. The calculated spin state splittings are correlated with experimental transition temperatures. The charge densities obtained with the B3LYP hybrid functional have been analysed in the framework of QTAIM.

< Table 1 >

**Results and Discussions**

The calculated spin state splittings depend sensitively on the method that has been used (Tab. 1). The Hartree-Fock method favours the HS state for all for complexes under study by about 300 kJ mol$^{-1}$. This behaviour has been observed earlier in other cases [6-8,10,13] and can be explained in a simple way: the Hartree-Fock methods neglects by definition any correlation between electrons with different spin polarization (Coulomb correlation) whereas the correlation between electrons with the same spin polarization (Fermi correlation) is partially taken into account due to the antisymmetry of the wave function. In case of the iron ion the electron-electron repulsion between the unpaired 3d electrons in the HS isomer is reduced by the Fermi correlation. Due to the neglect of Coulomb correlation there is no corresponding reduction of the electron-electron repulsion for the paired 3d electrons in the LS isomer. Consequently the calculated electronic energy for the LS state is too low in comparison to the energy of the HS state. The opposite behaviour is observed for the LDA method which favours the LS isomer by roughly 200 kJ mol$^{-1}$. Similar observations have been made for many other molecules but the reasons for this behaviour seem not to be clear [8].

The results obtained by the hybrid functionals B3LYP and B3LYP* and the GGA functionals BLYP, PBE, and BP86 exhibit (in this order) an increasing bias towards a LS ground state. The spin state splittings obtained with these functionals are in between the two extreme cases given by the Hartree-Fock and the LDA values, and they are obviously closer to the real value. The comparison of the calculated spin state splittings with experimental values for the complexes under study is hampered by two circumstances: the spin state splitting can be measured only indirectly (using the equilibrium constant or the transition temperature) and usually it is measured for solid samples. It has been observed that the type of counter ions sensitively influences the transition temperature [5] and hence also the spin state splitting. From the present data it is therefore not clear which method gives the most reliable value for the spin state splitting. Best agreement is obtained for the reparameterized B3LYP* functional. In order to take into account inter-molecular interactions, calculations with periodic boundary conditions – being computationally quite demanding – are currently performed and will be published elsewhere. Such calculations do not diminish the value of calculations for free molecules: only the combination of both types of calculation allows separating inter- and intra-molecular effects on the spin state splitting.

< Fig. 2 >

In many cases it is not the absolute value of the spin state splitting $E_S$ that is most interesting but the difference $\Delta E_S = E_S^{(\alpha)} - E_S^{(\beta)}$ between the spin state splittings of two molecules $\alpha$ and $\beta$. For the complexes under study it turns out that $\Delta E_S$ depends by far less sensitively on the calculational method than the absolute value $E_S$ does (Fig. 2). Comparing complex **1** with complexes **3** and **4** all methods give similar values for $\Delta E_S$. Comparing complex **1** with complex **2** the hybrid functionals and the GGA functionals give roughly the same values for $\Delta E_S$ within an error margin of 10 kJ mol$^{-1}$ while the values obtained with the Hartree-Fock and the LDA method deviate by about 20 kJ mol$^{-1}$. One may conclude that each of the calculational methods used here has a particular bias for the HS or LS state that is roughly constant for given class of molecules. Therefore, looking at the difference of spin state splitting between two similar molecules this bias should cancel to a large extent and any of the methods gives comparable results (excluding may be the two simplest ones, the Hartree-Fock and the LDA method). For completeness $\Delta E_S$ has been derived also from experimental results in a very approximate way by inserting an estimated entropy difference of 50 J mol$^{-1}$ into Eq. (4). In case of complex **2** the result is not in good agreement with the calculated values (Fig. 2) but it should be noted that the experimental values have been determined for solid samples which may easily exhibit a different spin state splitting than the free molecule.

These results illustrate that modern hybrid functionals or GGA functionals could be used to predict changes of the spin state splittings. The combination of calculations for free molecules and for the solid state [12] allows to separate inter- and intra-molecular effects and to improve the accuracy of the prediction. However, these methods do not have any 'explanatory power' if they are used as a 'black box' delivering only the spin state splitting as a number. At this point the QTAIM may be used to analyse the charge density obtained from DFT methods in order to gain a deeper understanding of calculated spin state splittings.

< Fig. 3 >

Next to the change of the spin state the most striking observation for SCO complexes is the drastic change of metal ligand bond lengths. The increase of bond length when going from the LS to the HS isomer should be accompanied by loosening of the bond. In the QTAIM framework a measure for the bond strengths is given by the negative Laplacian of the charge density at the bond critical point $\mathbf{r}_{bcp}$, $L(r_{bcp}) = -(\hbar^2/4m)\nabla^2\rho(r_{bcp})$. At the bond critical point the gradient of the charge density vanishes and the curvature of the charge density is positive in one direction (approximately the direction of the bond) and negative in two other mutually perpendicular directions. The function $L(\mathbf{r})$ indicates a local concentration ($L > 0$) or depletion ($L < 0$) of charge at point $\mathbf{r}$. $L(\mathbf{r})$ can be also interpreted as local energy density. Positive values (charge concentrations) indicate an excess of potential energy. X-ray diffraction experiments have confirmed that $L(\mathbf{r}_{bcp})$ decreases exponentially at the bond critical with increasing bond length [16]. Also calculations for a variety of SCO complexes with central [Fe(II)N6] octahedron (including the complexes under study) give an exponential relation between $L(\mathbf{r}_{bcp})$ and the iron ligand bond length (Fig. 3) and illustrate nicely the decrease of bond strength upon crossover from the LS to the HS state.

< Table 2 >

The QTAIM gives a rigorous definition of atomic charges in molecules. These atomic charges can of course be derived from the calculated charge density of the molecule and they can – although not easily – in principle be measured, and it has been shown that the charges of functional groups are with good accuracy the same in different molecules [14,15]. There exist a variety of chemical concepts that predict how the charges change when the molecule is changed. In order to utilize these concepts for the description of SCO features it is a tempting idea to correlate atomic charges of SCO complexes with the calculated spin state splittings. A very simple test to asses the accuracy of the calculated atomic charges is the comparison of the calculated total electronic charge with the true value which is of course known exactly. In case of complex **1** there are 248 electrons and the calculated electronic charge amounts to 247.9989 a.u. for the LS isomer and to 247.9816 a.u. for the HS isomer corresponding to relative errors of 4 and 74 ppm, respectively. For complexes **2** to **4**, which contain larger numbers of electrons, the respective relative errors are smaller. All complexes show roughly the same pattern what concerns the calculated atomic charges. In the LS isomers the iron centre, the apical carbon atom, and the pyrazol carbon atoms C1 and C3 are positively charged (+1.3, +0.9, +0.3, and +0.3 a.u., respectively) while the nitrogen atoms are negatively charged (−0.6 a.u.). All the remaining atoms are almost neutral (−0.1 to +0.15 a.u.). Roughly the same pattern can be observed for the HS isomers, where the largest changes are observed for the iron centre, which is more positive in the HS state (+1.5 a.u.). The differences of atomic charges between HS and LS isomers are quite small and the accuracy of the calculated total charge is slightly better for the LS isomers (see above). In order to avoid the calculated charge differences (HS-LS) to be biassed, all calculated atomic charges have been scaled in a way that the resulting total charge is correct (e.g. the calculated electronic charge of each atom of the HS isomer of complex **1** – that is 24.54 a.u. in case of iron – is multiplied by 1.000074 and all electronic charges of the LS isomer are multiplied by 1.000004, before the difference HS-LS is formed). The calculated differences (HS-LS) of atomic charges (Tab. 2) are surprisingly similar for all complexes studied here. Most relevant for the spin state splitting of these complexes are probably the charges of the iron centre and of the six coordinating nitrogen atoms (N1). In case of complexes **1**, **3**, and **4** the charge differences (as well as the calculated spin state splittings) are quite similar, whereas complex **2** exhibits a more negative charge difference for the six N1 and a more positive charge difference for the iron centre. In total in complex **2** there is upon crossover from the LS to the HS state a larger flow of charge from the coordinating nitrogen atoms to the central iron centre as compared to the other complexes. According to the results of all methods this larger charge flow is correlated to a stabilisation of the HS state (Tab. 1). Obviously this larger charge flow must be caused by the substitution of the hydrogen atom at position 1 with a methyl group. Substitution of the hydrogen atom at position 2 with either a methyl group (complex **3**) or with a bromine atom (complex **4**) does not have the same effect. The reason for this is not clear and the observed changes are small. At this point an extension of the investigation to a larger group of complexes will be necessary to be perfectly sure that the calculated numbers are significant.

< Table 3 >

The QTAIM allows also to break up the total electronic energy (including nuclear repulsion) into atomic contributions in the same rigorous way as it can be done for the total electronic charge. The results (Tab. 3) demonstrate that the influence of the substituent on the atomic energy of other atoms is quite small in comparison to the absolute values but large in comparison to the spin state splitting. The sum of the atomic contributions has a relative error of up to 200 ppm as compared to the total energy derived from the wave function. Similar numerical errors were observed for the atomic charges but in this case the problem is more serious since the spin state splitting amounts to about 10 ppm of the total energy. If the errors of the atomic energy contributions would be stochastic, these numerical errors would prevent to break up the spin state splitting into atomic contributions. However, one may assume that the numerical errors of the calculated atomic energies are largely proportional to the atomic energy itself. The following discussion of atomic energies is based on this assumption.

< Table 4 >

In order to overcome the numerical errors of the atomic energies, all atomic contributions to the total energy gained by the QTAIM analysis for a given complex and spin state are scaled by a fixed factor in such a way that the sum matches the total energy obtained directly from the wave function. Inspecting the resulting differences (HS-LS) of atomic energies (Tab. 4) reveals large changes for almost all atoms and functional groups (R1 and R3 being the only exceptions). It should be noted that the sum of atomic contributions for a complex in Table 4 (each contribution weighted according to the molecular stoichiometry) does not exactly match the corresponding spin state splitting calculated with B3LYP (Tab. 1) due to rounding effects. The changes between complex **1** and **3** on one side and complex **2** on the other side are mostly due to changes within the first coordination sphere of the iron centre. In complex **2** the increase of atomic energy of the iron centre upon spin crossover is clearly larger than in complexes **1** and **3** (131 kJ/mol as compared to 54 and 77 kJ/mol, respectively). But this effect is more than compensated by decreased energy gain of the six nitrogen atoms N1 (6 x 23 kJ/mol as compared to 6 x 67 and 6 x 68 kJ/mol, respectively). The results for complex **4** do not fit into this picture. A possible explanation may be that the linear scaling of the atomic energies is not appropriate in this case due to the numerical errors of the large atomic energy contributions of the bromine atom ($\approx 7 \cdot 10^6$ kJ/mol).

Summarizing the results of the QTAIM analysis the following simplified picture may used to explain the different spin state splittings for the complexes under study: Substitutions on position 2 with a methyl group or a bromine atom do not have a significant effect on the spin state splitting. Substitution of the hydrogen atom on position 1 (complex **2**) instead leads to an increased flow of electronic charge (partially from the iron centre) to the nitrogen atoms N1 upon crossover from the LS to the HS state, accompanied by an increased stabilisation of the nitrogen atoms and of the molecule as a whole in the HS state. It will be left to future investigations to test the validity of this picture with a larger class of molecules.

**Materials and Methods**

Electronic structure calculations have been performed for free molecules with a variety of different methods, namely the local density approximation (LDA, using the functional V from Vosko, Wilk, and Nusair [17], which fits the Ceperly-Alder solution to the uniform electron gas), the generalized gradient approximation (GGA) functionals BLYP (Becke's exchange functional [18] with the correlation functional of Lee, Yang, and Parr [19,20]), PBE (the exchange and correlation functional from Perdew, Burke, and Ernzerhof [21,22]), and BP86 (Becke's exchange functional [18] together with Perdew's correlation functional [23]), the hybrid functionals B3LYP [24] and B3LYP* [10], and the Hartree-Fock method (HF). The 6-311G(2d,p) basis for H, C, and N and the Wachters-Hay

double-$\zeta$ basis for Fe [25,26] (6-311G for short) have been used. All calculations were performed with the program package Gaussian 03 [27]. The spin state splitting $E_S$ for HS and LS states was calculated after full geometry optimisation for the respective spin isomers. The QTAIM analysis of the calculated charge densities has been done with the program AIM2000 [28].

< Scheme 1 >

The calculations have been performed for four different iron complexes with different substituents. The mother complex is formed by an iron(II) centre and two tris(pyrazol-1-yl)methane (tpm) ligands (complex **1**, [Fe(tpm)$_2$]$^{2+}$, Fig. 1) [29]. Three more complexes can be obtained by substituting hydrogen atoms of the pyrazol rings (Scheme 1) with methyl groups or bromine [5]: [Fe(t3mpm)$_2$]$^{2+}$ (complex **2**, t3mpm = tris(3-methylpyrazol-1-yl)methane), [Fe(t4mpm)$_2$]$^{2+}$ (complex **3**, t4mpm = tris(4-methyl-pyrazol-1-yl)methane), [Fe(t4bpm)$_2$]$^{2+}$ (complex **4**, t4bpm = tris(4-bromo-pyrazol-1-yl)methane). The spin state splitting for complexes **1** and **2** obtained with the B3LYP and BLYP functional and for complex **1** with the HF method are taken from Ref. [5].

**Acknowledgments**

Financial support by the Deutsche Forschungsgemeinschaft (DFG) is gratefully acknowledged (Tr 97/31 and Tr 97/32). We thank J.-P. Tuchagues and H. Winkler for fruitful discussions.

*Figure Captions*

**Fig. 1.** Molecular structure of the HS isomer of complex **1** derived from geometry optimization with B3LYP/6-311G. Hydrogen atoms have been omitted for clarity

**Fig. 2.** Differences of spin state splitting $\Delta E_S = E_S^{(\alpha)} - E_S^{(1)}$ between complex $\alpha$ and complex **1** (□: $\alpha = 2$, ▲: $\alpha = 3$, ●: $\alpha = 4$) obtained by various methods. Boxes with sparse hatching illustrate the variances for GGA and hybrid functionals

**Fig. 3.** Negative Laplacian L(**r**) at the Fe-N bond critical points of a variety of iron(II) SCO complexes and of $[Fe(CN)_5(NO)]^{2-}$ (▲: LS, ●: HS). Values are obtained with AIM2000 [28] from charge densities calculated with B3LYP/6-311G

**Scheme 1.** Drawing of the tris(pyrazol-1-yl)methane ligands tpm ($R_1=R_2=R_3=H$), t3mpm ($R_1=CH_3$, $R_2=R_3=H$), t4mpm ($R_1=R_3=H$, $R_2=CH_3$), and t4bpm ($R_1=R_3=H$, $R_2=Br$)

**Table 1**. Spin state splittings $E_S$ (kJ mol$^{-1}$) estimated according to experimental transition temperatures [5] and calculated with different methods as indicated

|  | Complex | | | |
|---|---|---|---|---|
| Method | 1 | 2 | 3 | 4 |
| Experiment | 18[a] | ≈ 10[a,b] | ≈ 20[a,b] | ≈ 20[a,b] |
| HF | -300 | -319 | -300 | -301 |
| B3LYP | -7 | -40 | -7 | -8 |
| B3LYP* | 21 | -13 | 22 | 20 |
| BLYP | 82 | 39 | 83 | 79 |
| PBE | 103 | 62 | 103 | 99 |
| BP86 | 112 | 72 |  | 109 |
| LDA | 234 | 208 |  | 230 |

[a] estimated according to $T_{1/2}$ taken from [29] and assuming $\Delta S \approx$ 50 J mol$^{-1}$ K$^{-1}$; [b] estimated according to $T_{1/2}$ taken from [5]

**Table 2**. Differences (HS-LS) of QTAIM atomic charges (a.u.) for complexes **1** to **4**. Atoms and functional groups are numbered according to Scheme 1

| Atom | 1 | 2 | 3 | 4 |
|---|---|---|---|---|
| Fe | 0.176 | 0.186 | 0.176 | 0.183 |
| N1 | -0.018 | -0.027 | -0.020 | -0.020 |
| N2 | 0.012 | 0.011 | 0.010 | 0.008 |
| C1 | -0.001 | 0.000 | -0.001 | 0.000 |
| C2 | -0.002 | -0.001 | -0.002 | -0.002 |
| C3 | -0.003 | -0.001 | -0.002 | -0.002 |
| $C_{apical}$ | -0.008 | -0.007 | -0.007 | -0.004 |
| $H_{apical}$ | -0.011 | -0.008 | -0.010 | -0.010 |
| R1 | -0.005 | -0.004 | -0.004 | -0.004 |
| R2 | -0.002 | -0.001 | -0.002 | -0.003 |
| R3 | -0.003 | -0.004 | -0.003 | -0.002 |

**Table 3**. QTAIM atomic energies ($10^6$ J/mol) for selected atoms or functional groups (atom numbering according to Scheme 1) for the HS isomers of complexes **1** to **4**

| Atom | 1 | 2 | 3 | 4 |
|---|---|---|---|---|
| Fe | -3319.3 | -3319.7 | -3319.9 | -3316.8 |
| N1 | -144.2 | -144.2 | -144.2 | -144.1 |
| N2 | -144.7 | -144.7 | -144.7 | -144.6 |
| R1 | -1.5 | -1.5 | -1.5 | -1.5 |
| R2 | -1.5 | -1.5 | -104.4 | -6763.3 |
| R3 | -1.5 | -104.5 | -1.5 | -1.5 |

**Table 4**. Differences (HS-LS) of QTAIM atomic energies (kJ/mol) for complexes **1** to **4**. Atoms and functional groups are numbered according to Scheme 1

| Atom | 1 | 2 | 3 | 4 |
|---|---|---|---|---|
| Fe | 54 | 131 | 77 | 434 |
| N1 | 67 | 23 | 68 | 83 |
| N2 | 15 | 24 | 18 | 29 |
| C1 | -27 | -22 | -25 | -13 |
| C2 | -20 | -11 | -20 | -9 |
| C3 | -29 | -34 | -28 | -19 |
| C$_{apical}$ | -8 | -1 | -6 | 2 |
| H$_{apical}$ | -18 | -11 | -17 | -18 |
| R1 | -4 | -4 | -4 | -4 |
| R2 | -1 | 0 | -14 | -135 |
| R3 | -1 | 0 | -1 | -1 |

Fig. 1

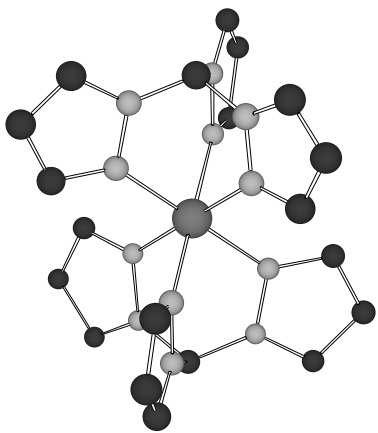

Fig. 2

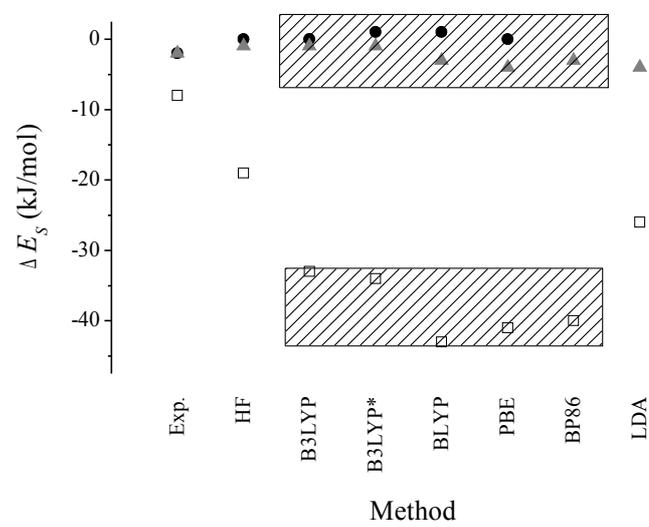

Fig. 3

L(r) (e/Å³) vs r(Fe,N) (Å)

Scheme 1

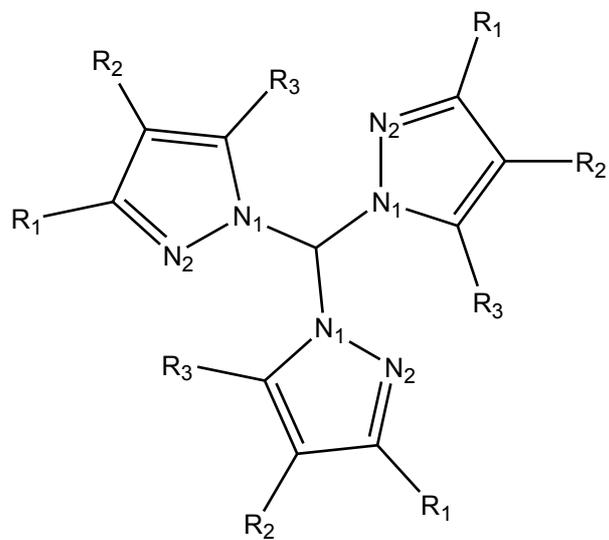

# Graphics for use in the Table of Contents

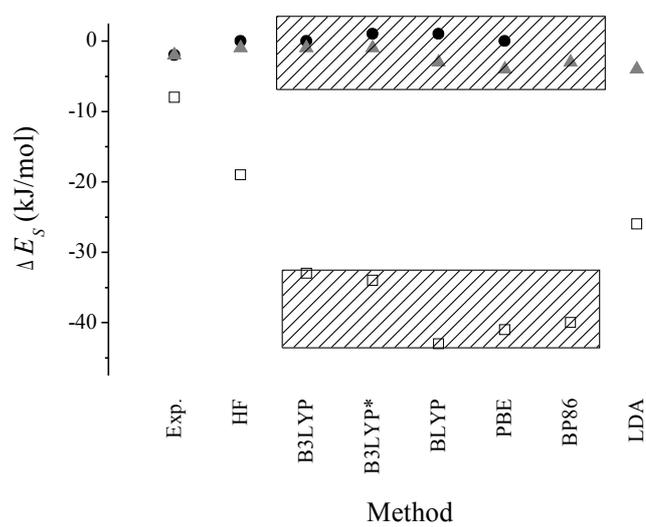